\documentclass[twocolumn,showpacs]{revtex4} 

\usepackage{color}
\usepackage{graphicx}

\newcommand{\etal}{{\it et al.\/}\ }

\begin{document}

\title{Electronic structure of solid coronene: differences and commonalities to picene}
\author{Taichi Kosugi$^{1}$}
\author{Takashi Miyake$^{1,2}$}
\author{Shoji Ishibashi$^1$}
\author{Ryotaro Arita$^{2,3,4}$}
\author{Hideo Aoki$^{5}$}

\affiliation{$^1$Nanosystem Research Institute ``RICS", AIST, Umezono, Tsukuba 305-8568, Japan}
\affiliation{$^2$ Japan Science and Technology Agency, CREST, Honcho, Kawaguchi, Saitama 332-0012, Japan}
\affiliation{$^3$ Department of Applied Physics, University of Tokyo, Hongo, Tokyo 113-8656, Japan}
\affiliation{$^4$ PRESTO, Japan Science and Technology Agency (JST), Kawaguchi, Saitama 332-0012, Japan}
\affiliation{$^5$ Department of Physics, University of Tokyo, Hongo, Tokyo 113-0033, Japan}

\begin{abstract}
We have obtained the first-principles electronic structure of solid coronene, 
which has been recently discovered to exhibit superconductivity with potassium doping.
Since coronene, along with picene, the first aromatic superconductor, now provide 
a class of superconductors as solids of aromatic compounds, here 
we compare the two cases in examining the electronic structures.  
In the undoped coronene crystal, where the molecules are arranged in a 
herringbone structure with two molecules in a unit cell, 
the conduction band above an insulating gap 
is found to comprise four bands, which 
basically originate from the lowest two unoccupied molecular orbitals 
 (doubly-degenerate, reflecting the high symmetry of the molecular shape) in an isolated molecule but the bands are 
entangled as in solid picene.   The Fermi surface for a candidate 
of the structure of K$_x$coronene with $x=3$, for 
which superconductivity is found,  comprises multiple sheets, 
as in doped picene but exhibiting a larger anisotropy with different topology.
\end{abstract}


\maketitle

{\it Introduction ---} 
Discovery of superconductivity in solid picene doped with potassium 
atoms~\cite{kubozono} is seminal and came as a surprise 
in that the picene, on top of 
the highest $T_c$ among organic superconductors, 
belongs to 
{\it aromatic} compounds, the most typical, textbook 
class of organic materials.  
Then a natural question to ask is: can other aromatic compounds 
become superconducting as well?  
Before examining this question, let us first briefly summarize the known 
superconductors in carbon-based materials.  
First discovery goes back to 1965, when a 
graphite-intercalation compound (GIC), KC$_8$, 
was found to be a superconductor at 0.1 K \cite{bib:1114}, 
and the highest $T_c$ among GIC's to date is 11.6 K in CaC$_6$\cite{bib:1108_10-11}. 
Fullerene is another class, where 
potassium-doped fullerene, K$_3$C$_{60}$, has 
$T_c = 20$ K \cite{bib:1104_1}, followed by 
Cs$_2$RbC$_{60}$ with $T_c = 33$ K~\cite{bib:1104_4}, 
or 40 K in Cs$_3$C$_{60}$ under 15 kbar \cite{bib:1104_8}. 
In the last decade, the boron-doped diamond \cite{ekimov} joined 
carbon-based superconductors. 
The first aromatic superconductivity in the 
potassium-doped solid picene discovered by 
Kubozono's group\cite{kubozono}, 
with $T_c=7-18$ K in K$_x$picene  for $x\simeq 3$, 
was unsuspected, since 
the (undoped) solid of picene (a hydrocarbon compound with 
five benzene rings connected in a zigzag) has been known to 
be a good insulator, as naturally expected for an aromatic molecule.   

The present authors have reported for the first time 
the first-principles electronic 
structure of both undoped and doped solid picene in the framework of the density functional theory (DFT) within the local density approximation 
(LDA) \cite{kosugi}.  
 We have revealed, first for the undoped solid picene, 
where picene molecules take a 
herringbone structure, that the conduction band 
consists of four bands, which originate basically 
 from the lowest two unoccupied molecular orbitals (LUMO and LUMO+1) 
of an isolated molecule, but the bands are entangled ( i.e., crossing with each other.   When doped with potassium atoms, 
the herringbone structure is deformed, and the electronic 
wave function significantly spills from the organic molecules' LUMO's 
into potassium sites\cite{kosugi}. 

Now, if we come back to the question of whether and which other aromatic 
compounds can become superconducting, recently Kubozono's group 
reported superconductivity in doped coronene \cite{kubozono2}. 
Coronene, C$_{24}$H$_{12}$, is another of typical aromatic molecules, 
with seven benzene rings assembled in a concentric disk.  
Superconductivity in K$_x$coronene is reported to 
appear around $x\simeq 3$ with $T_c$ up to 15 K.\cite{Wang}
Motivated by this, here we have obtained the electronic structure of 
solid coronene, both undoped and doped.  
An obvious interest is the differences and commonalities 
between the crystal and electronic structures of 
solid coronene as compared with those of solid picene.  
This is precisely the purpose of the present work.   
The obtained results are analyzed with maximally-localized
Wannier functions (WF's)~\cite{bib:MLWF}, in terms of which 
we downfold the system into 
a tight-binding model and 
compare with those for picene.

We shall show that 
the conduction band of the undoped solid coronene 
comprises four bands, which 
basically originate from the two LUMO's 
 (doubly-degenerate, reflecting the symmetry of the molecule higher than 
that of picene) in an isolated molecule, but the bands are 
entangled as in solid picene.
The Fermi surface for a candidate 
of the structure of K$_x$coronene with $x=3$, for 
which superconductivity is found,  comprises multiple sheets, 
as in doped picene but exhibiting a more one-dimensional 
character with different anisotropy nad topology.

The present electronic structure will serve as a basis for discussing mechanisms of the superconductivity.  
Since some classes of organic superconductors are considered 
to have an electronic mechanism,\cite{kurokireview} picene and coronene 
superconductors may possibly belong to them.  
There have been some theoretical studies\cite{Giovannetti,Kim} 
that suggest that solid
picene is a strongly correlated electron system.  
As for the electron-phonon coupling, on the other hand, 
Kato {\it et al.} estimated the phonon frequencies and the electron-phonon 
coupling for various hydrocarbon molecules\cite{Kato1}.
For picene 
they found that the electron-phonon coupling 
can be as large as $0.2$ eV.  
While it is generally difficult to quantitatively estimate $T_c$ 
for a phonon mechanism 
due to the ambiguity in the Coulomb pseudopotential ($\mu^*$), they concluded that $T_c \sim 10$ K may be expected.
They also estimated the coupling for coronene to be about $0.1$ eV.
Subedi {\it et al.}\cite{Subedi} recently calculated
the phonon spectrum and electron-phonon interaction for solid picene,
and they found that
the calculated electron-phonon coupling $0.1$-$0.2$ eV is sufficiently strong
to reproduce the experimental $T_c$ of 18 K within the Migdal-Eliashberg
theory.

{\it First-principles bands ---} 
The calculation is based on DFT, 
where LDA in the Perdew-Zunger formula 
is adopted for the exchange-correlation energy functional.\cite{bib:LDA}
We use the projector-augmented wave (PAW) method\cite{bib:PAW}, implemented to 
the Quantum MAterials Simulator (QMAS) package.\cite{bib:QMAS}
The pseudo Bloch wave functions are expanded by plane waves 
up to an energy cutoff of 40 Ry with $4 \times 6 \times 4$ $k$-points.

\begin{figure}[htbp]
\begin{center}
\includegraphics[keepaspectratio,width=5.5cm]{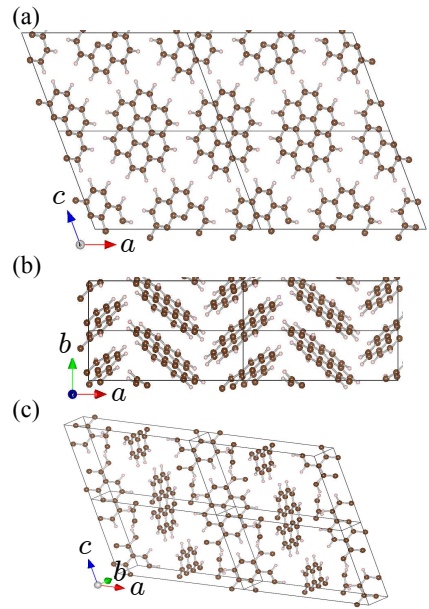}
\end{center}
\caption{
(Color online) 
Crystal structure of undoped solid coronene viewed along (a) $b$ axis, (b) $c$ axis and (c) a bird's eye view. Solid lines delineate unit cells.
}
\label{crststr}
\end{figure}

As for the crystal structure, here we adopt the experimental lattice parameters of natural coronene (karpatite) reported by Echigo \etal\cite{bib:1608}
Natural coronene has a monoclinic (space group: $P2_1/a$) 
structure with $a=16.094, b=4.690$ and $c=10.049$ \AA, and $\beta=110.79^\circ$.  
We have also performed calculations adopting 
the lattice parameters of synthetic coronene~\cite{bib:1606}, 
and have confirmed that the results are essentially unchanged 
from those for the natural coronene.
The molecular solid has a herringbone arrangement of molecules as depicted in Fig. \ref{crststr} with a unit cell containing two molecules 
(centered respectively at $(0,0,0)$ and $(1/2,1/2,0)$ as 
dictated by the symmetry).
The lattice parameters are fixed at the experimental values, and the internal atomic positions are optimized. 
The angle between the planes of the inequivalent molecules in the optimized geometry is 
$95^\circ$, which agrees with the measured value.  
It was found that the point-group symmetry $D_{6h} $ for an isolated coronene molecule is lowered to $D_{2h}$ in the crystal.

Figure 2 displays the electronic band structure of coronene.
We have an insulator with a band gap of $2.41$ eV.   
The gap, which is naturally 
smaller than the HOMO-LUMO gap, calculated to be $2.90$ eV, of an isolated coronene molecule,  is indirect, with 
the valence band top located at Y in the Brillouin zone, 
while the conduction band bottom at $\Gamma$. 
Since the crystal structure is layered, where the herringbone 
arrangement, on the $a-b$ plane, of the molecules are stacked along $c$ axis, 
the electronic structure is more dispersive 
along $a^*-b^*$ axis.  More precisely, the dispersion 
along  $b^*$ is larger than that for $a^*$, which directly 
reflects the distance between neighboring molecules being 
shortest along $b$.
The conduction band, with a width of $0.40$ eV, consists of four bands derived from the doubly-degenerate, $e_{1g}$ LUMO's.  
For solid picene by comparison,  the conduction band, 
with a width $0.39$ eV, consists of four bands with a band gap of $2.36$ eV, while the LUMO-HOMO gap of a picene molecule is $2.96$ eV. 
These are similar to solid coronene, where a difference is the conduction 
band of the solid picene originates from LUMO and LUMO+1 of an isolated picene molecule \cite{kosugi}.
Considering the general tendency of LDA to underestimate band gaps, 
we expect that the actual band gap of coronene may be larger than the calculated value.  
An accurate prediction of optical properties and band widths will also require incorporation of many-body effects, 
as is done with GW by Roth \etal\cite{bib:1574} for picene.
The valence band of solid coronene has a width of $0.45$ eV, and 
consists of four bands,
which is shown to be derived from the (again doubly-degenerate) 
$e_{2u}$ HOMO's of an isolated coronene molecule.

\begin{figure}[htbp]
\begin{center}
\includegraphics[keepaspectratio,width=8cm]{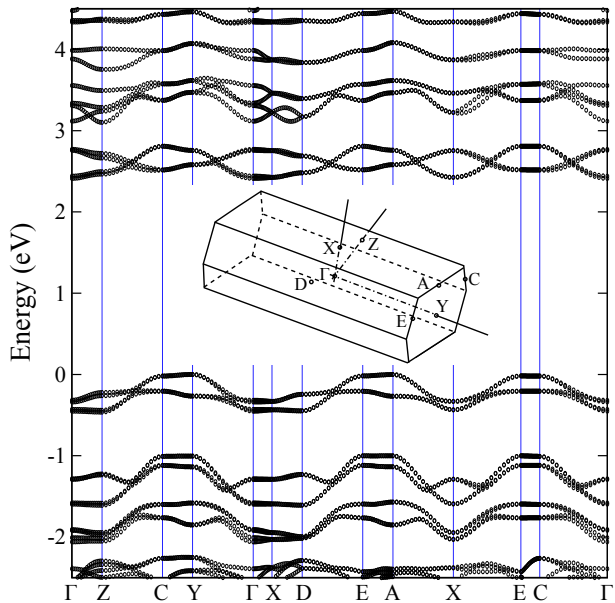}
\end{center}
\caption{
(Color online) 
Calculated electronic band structure of the undoped crystalline coronene.
The origin of the energy is set to be the valence band top.
The inset depicts the Brillouin zone with 
$\Gamma$, Z, C, Y, X, D, E and A respectively corresponding to 
$(0,0,0)$, $(0,0,1/2)$, $(0,1/2,1/2)$, $(0, 1/2, 0)$, $(1/2, 0, 0)$, $(1/2, 0, 1/2)$, $(1/2, 1/2, 1/2)$ and $(1/2, 1/2, 0)$ in units of $(\mathbf{a}^*, \mathbf{b}^*, \mathbf{c}^*)$.  
}
\label{band1}
\end{figure}

\begin{figure}[htbp]
\begin{center}
\includegraphics[keepaspectratio,width=8cm]{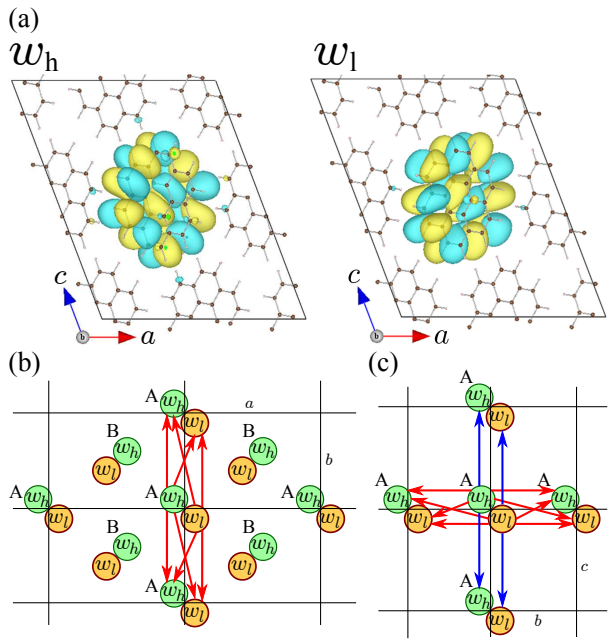}
\end{center}
\caption{
(Color online) 
(a) displays maximally localized Wannier functions (right) $w_l$  ($w_h$) 
with a lower-energy (higher-energy) 
constructed from the conduction bands of undoped solid coronene. 
Their transfer integrals $t$ on the $a-b$ plane (b) and $b-c$ plane (c) are also shown, where 
the red (blue) arrows are for $|t| > 30$ meV ($10 < |t| < 30$ meV).
}
\label{wannier}
\end{figure}

{\it Downfolding ---}
We have constructed an \textit{ab initio} tight-binding model from the maximally localized WF's~\cite{bib:MLWF} of the four-fold conduction band.  
The two-fold degenerate $e_{1g}$ LUMO's for an isolated molecule have $b_{3g}$ and $b_{2g}$ symmetries in the $D_{2h}$ representation.
Figure \ref{wannier}(a) displays the two WF's localized at each of the molecules.  
While the downfolded tight-binding band (not shown) accurately reproduces the DFT-LDA band structure, we 
can notice that the WF $w_{\mathrm{h}}$ is derived mainly from $b_{3g}$ orbital while $w_{\mathrm{l}}$ from $b_{2g}$.  
The difference in the orbital energy between the two WF's is $18.2$ meV, which is to be compared with $11.3$ meV for solid picene.
Figure \ref{wannier}(b,c) depicts the major transfer integrals in the downfolded tight-binding model,
which exhibit the transfers along $b$ axis much stronger than the other axes,
in contrast to those in picene (see Fig. 3(c) in Ref.\cite{kosugi}),
for which two transfers on $a-b$ plane are close to each other.  
The anisotropy of the transfers in coronene thus comes from the 
relative distances between the WF's.

\begin{figure}[htbp]
\begin{center}
\includegraphics[keepaspectratio,width=8cm]{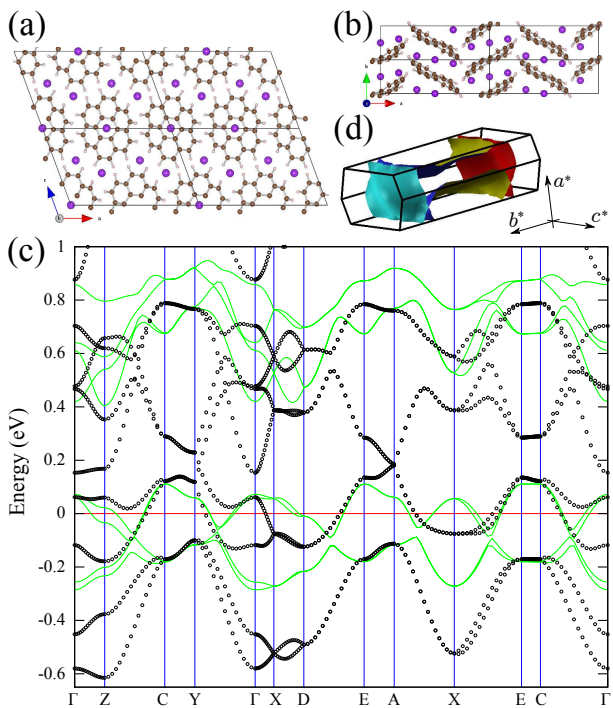}
\end{center}
\caption{
(Color online)
Crystal structure of K$_3$coronene viewed along (a) $b$ or (b) $c$ axis. 
Larger balls represent K atoms.
(c) shows electronic band structures of the K$_3$coronene (circles) and undoped coronene (curves).
The origin of energy for the K$_3$coronene is set to its Fermi level
while  that for undoped coronene set for $x=3$ in a rigid-band shift for 
comparison.  
The Fermi surface for the K$_3$coronene is displayed in (d).
}
\label{K3coronene}
\end{figure}

{\it Doped solid coronene ---} 
Let us finally discuss the doped coronene.  
If we naively adopt a rigid-band picture for a doping level of 
$x=3$ for which superconductivity is observed, 
the Fermi surface (not shown) consists of multiple surfaces that comprise 
one-dimensional (planar) surfaces
and a two-dimensional (cylindrical) one, along with a more three-dimensional one.  
However, since we have shown for the doped picene\cite{kosugi} 
that the rigid-band picture is broken in this material for two-fold reasons 
(i.e., a distorted herringbone structure upon doping along 
with a spilling of the molecular wave function over to 
potassium sites), we should have a look at the electronic 
structure of the coronene actually doped with K.  
Since the structure of the doped system has not been 
experimentally obtained, we have tried a structural optimization of K$_3$coronene.  Since this in itself is an important but vast task, 
here we only show a candidate structure.
We  started from a plausible geometry with one potassium atom just above the molecule and two in the interstitial region.   
The initial coordinates of the six potassium atoms we adopted are 
$(1/2, 0, 0), (0, 1/2, 0), (\pm 1/3, 1/2, \pm 1/3)$ and $(\pm 5/6, 0, \pm 1/3)$,
which preserve the monoclinic symmetry.
Interestingly, the optimization, with the lattice parameters fixed at those for the pristine structure, resulted in a significant rearrangement 
of the herringbone structure with the monoclinic symmetry lowered, 
along with a strong deformation of each molecule, 
as seen in Figs. \ref{K3coronene}(a) and (b).
The band structure of the K$_3$coronene is significantly more dispersive than the undoped one, where 
the LUMO-derived band group is fused with the upper band group, resulting in a much wider band group, as shown in Fig. \ref{K3coronene}(c).
We can see that the resultant Fermi surface is an intriguing composite of one-dimensional surfaces (i.e., multiple pairs of planar surfaces) with 
different anisotropies.  
The doped picene also has a coexistence of multiple Fermi surfaces\cite{kosugi}, but the doped coronene has topology of the surface 
different from those in doped picene, which comes from 
different (and more one-dimensional) 
tight-binding structure (Fig. \ref{wannier}(b,c)).   
The anisotropic, multiple Fermi surface 
should give a basis for examining superconductivity.   
In a phonon mechanism the problem becomes the coupling 
between the electrons on such a Fermi surface and the molecular 
phonons.  The situation gives an interesting possibility for electron 
mechanisms as well, since the nesting between disconnected Fermi 
surfaces can give rise to a unique opportunity for an electron 
mechanism, especially in multiband cases.\cite{aoki09}
For an accurate description of the structure more elaborate and exhaustive structural optimizations will be needed, 
since the large degrees of freedom on the dopant positions will render the energy surface many local minima.  
Such a systematic examination is, however, beyond the scope of the present work, and will be reported in future.

Finally, a few words about the ``aromaticity" of molecules.
Its simplest definition is in terms of the ``Clar sextets" (resonating benzene rings)\cite{clar}
on a given molecular structure.
When a molecule is fully benzenoid (where the sextets exhaust all the double bonds)
the wave function tends to be localized on the sextets.
In this picture picene is non-fully-benzenoid,
while coronene, also non-fully-benzenoid, has multiple Clar structures.\cite{bib:1606}
The relation of these quantum chemical properties with band structures is another of interesting future problems.

In summary, 
we have studied the electronic structure of solid coronene by means of first-principles calculation.  
The conduction band is found to comprise four bands, which 
basically originate from the lowest two unoccupied molecular orbitals 
 (doubly-degenerate, reflecting the molecular symmetry) in an isolated molecule, but the bands are 
entangled.  The 
maximally localized WF's are used to derive a 
downfolded tight-binding Hamiltonian, where the major 
transfer integrals are found to significantly differ from those in solid picene.  
 The Fermi surface for a candidate 
of the structure of K$_x$coronene with $x=3$, for 
which superconductivity is found,  comprises multiple sheets, 
as in doped picene but with different topology of the surface.
Their relevance to superconductivity is an interesting future problem.

We are indebted to Yoshihiro 
Kubozono for letting us know of the experimental 
results prior to publication.  
The present work is partially supported by the Next Generation Supercomputer Project,
Nanoscience Program from MEXT, Japan, and 
by Grants-in-aid No. 19051016 and 22104010 from MEXT, Japan and the JST PRESTO program.
The calculations were performed at the supercomputer centers of ISSP, 
University of Tokyo, and at the Information Technology Center, University of Tokyo.

\end{document}